\documentclass[a4paper,12pt]{article}

\usepackage[center]{titlesec}

\renewcommand{\thesection}{\Roman{section}}

\renewcommand{\thesubsection}{\arabic{subsection}}

\renewcommand{\thesubsubsection}{\alph{subsubsection}}

\titleformat%
    {\section}
    [block]
    {\centering\bfseries}
    {\thesection.}
    {0.5em}
    {}

\titleformat%
    {\subsection}
    [block]
    {\centering\itshape}
    {\thesubsection.}
    {0.5em}
    {}

\titleformat%
    {\subsubsection}
    [block]
    {\centering}
    {\thesubsubsection.)}
    {0.5em}
    {}

\usepackage{lipsum}

\titlespacing{\section}{0pt}{*4}{*1.5}
\titlespacing{\subsection}{0pt}{*4}{*1.5}
\titlespacing{\subsubsection}{0pt}{*4}{*1.5}

\usepackage[utf8]{inputenc}
\usepackage{cancel}
\usepackage{tikz}
\usepackage{ulem}
\usepackage{amsfonts}
\usepackage{amssymb}
\usepackage{graphicx}
\usepackage{amsmath}
\usepackage{enumerate}
\usepackage{mathtools}
\usepackage{subfig}
\usepackage{color}
\usepackage{tikz}
\usepackage{float}
\usepackage{here}
\usepackage{cite}
\usepackage{mathrsfs}
\usepackage{float,epsfig}
\usepackage{dcolumn}
\usepackage{graphicx}
\usepackage{bm}
\usepackage{amsmath,amssymb,amsthm}
\usepackage[colorlinks=true,linkcolor=blue,citecolor=red]{hyperref}
\usepackage{multirow}
\usepackage[toc,page]{appendix}
\usetikzlibrary{arrows.meta}
\usetikzlibrary{bending}
\usetikzlibrary{calc}
\newcommand{\be}{\begin{equation}}
\newcommand{\ee}{\end{equation}}
\newcommand{\bea}{\setlength\arraycolsep{2pt} \begin{eqnarray}}
\newcommand{\eea}{\end{eqnarray}}

\def\0{{\sst{(0)}}}
\def\1{{\sst{(1)}}}
\def\2{{\sst{(2)}}}
\def\3{{\sst{(3)}}}
\def\4{{\sst{(4)}}}
\def\5{{\sst{(5)}}}
\def\6{{\sst{(6)}}}
\def\7{{\sst{(7)}}}
\def\8{{\sst{(8)}}}
\def\sst#1{{\scriptscriptstyle #1}}

\makeatletter \@addtoreset{equation}{section}

\definecolor{lime}{HTML}{A6CE39}

\begin{document}

\title{{\normalsize \textbf{\Large    Constrained     Deflection Angle  and  Shadows  of   Rotating Black Holes in  Einstein-Maxwell-scalar
Theory }}}
\author{ {  Hajar Belmahi\thanks{hajar-belmahi@um5.ac.ma} \hspace*{-8pt}} \\
{\small Faculté des Sciences, Université Mohammed V de Rabat, Rabat, Morocco}}
\maketitle

\begin{abstract}
Motivated by recent Event Horizon Telescope  findings, we investigate   constrained   optics of rotating black holes in  Einstein-Maxwell-scalar gravity theory. Precisely, we mainly study the  parameter effects on  two relevant optical concepts being the shadow and deflection angle. Using the Hamilton-Jacobi algorithm, we find certain shadow geometries being corroborated by observational data once appropriate constraints on a stringy coupling  parameter $\beta$  are imposed. For such   constrained  regions of  the black hole  moduli space, we compute    the deflection angle of light rays near to such black holes.  Concretely,   we show that this optical quantity  can be split into  two  contributions describing the absence and the presence of the coupling  parameter $\beta$.  Then, we discuss and  analyse such  an optical quantity  in  terms of     $\beta$  contributions. 

\end{abstract}


\section{Introduction}

Recently, the study of the optical aspect of black holes becomes a warm subject being largely investigated using different approaches including analytical and numerical results brought either by Event Horizon Telescope (EHT) collaborations or by the gravitational wave detections \cite{A56,B10,A57,A58,A59,A60,LIGs,BGRA}. A close inspection reveals that two relevant concepts have been studied being the shadow and the deflection angle of the light rays near to several black holes \cite{Jaf,Yanxx,A154A,A65,A66,A68,A69,A74,A78,A81,MHER,HERDE,Novo,SV1}. A particular emphasis has been put on  such physical proprieties of black holes with non-trivial backgrounds. In this way, the shadow aspect, providing predictions that can be analyzed using the falsification mechanism, has been largely studied via  various techniques. Exploiting the Hamilton-Jacobi formalism, certain geometrical configurations of black hole shadows have been obtained and analyzed. More precisely, regular and irregular black holes, in many gravity theories, have been approached.  Concretely,  it has been shown that the shadows of the non-rotating black holes
exhibit   circle configurations  with different sizes  controlled by  the black hole  parameters including 
 the charge.  However, such perfect geometries have been deformed by  introducing the rotating parameter providing 
non-trivial one dimensional curves  known as D or cardioid ones\cite{A79,A159,SV2,A77}.

Motivated by high-energy physical models such as string theory, the shadows of black holes in the  Starobinsky-Bel-Robinson (SBR) gravity by introducing a stringy parameter have been investigated  \cite{sbrh}. It has been revealed that this parameter can modify previous  shadow results.  Alternatively, other investigations  relying on different stringy gravities   have been also conducted. Among others,    it has been provided various shadow geometries, aligning with EHT collaboration data through imposed  constraints on the involved black hole parameters \cite{Lekbi,YASSS,Tsu,BBAM,AMM}. 
 
Moreover, many efforts have been devoted to study light behaviors near to certain  anti de Sitter (AdS) black holes.  This concerns the  computation  and the  analysis of the deflection angle of light rays near to various black holes in different models, including those related to high-energy physics \cite{A95,A89,A96}. For  the AdS black holes, this optical quantity for both regular and non-regular black holes, with  the presence or the absence of dark energy sectors, has been approached via graphical discussions by varying the relevant involved parameters. In particular, the deflection angle of light rays near   to four and seven-dimensional AdS black holes, derived from M-theory compactifications,  has been computed by unveiling the impact of the M-brane number coupled to  the rotating parameter \cite{HRC,HADS,HMTD}.

More recently, a special interest  has been put on the study of black holes in modified  Einstein gravity by implementing a possible coupling between 
 scalar fields  and  gauge fields.   In string theory, for instance,  these scalar fields can be obtained  from  either the stringy  spectrum  including the dilaton and  the  axion or the compactification scenarios.  Thermodynamic and optical behaviors of such black holes have been largely investigated  furnishing  interesting results \cite{Hirschmann, Yu, Al, Charmousis,Lu,hemd}.

The aim of this work is  to   provide   constrained   optics of rotating black holes in  Einstein-Maxwell-scalar (EMS) gravity theory. Precisely, we  investigate certain   parameter effects on  two relevant optical concepts being the shadow and the  deflection angle. Using the Hamilton- Jacobi algorithm, we find certain shadow geometries being corroborated by observational data by imposing  appropriate constraints on a stringy coupling  parameter $\beta$. For such   constrained  regions of  the black hole  moduli space, we study  the deflection angle of light rays near to such black holes.  Precisely,   we  reveal  that this optical quantity  can be split into  two  contributions describing the absence and the presence of the coupling  parameter $\beta$.  Then, we discuss and  analyse such  an optical quantity  in  terms of    $\beta$  contributions.

This work is structured   as follows. In section II, we  provide  a concise discussion on 
prorating  black holes in  EMS theory in the presence of a stringy coupling parameter $\beta$.  In section III,  we  investigate the  associated shadows. In section IV,  we examine  the deflection angle  in terms of the stringy coupling parameter. Section V  provides concluding discussions.

\section{Rotating EMS black holes}
In this section, we expose the black hole models in EMS gravity theory. This theory 
has been supported by superstring  model scenarios where various scalar fields appear in
natural ways, belonging  to lower energy limits of supergravity in
different dimensions. These scalars including the dilaton and the axion could couple to
other spectrum gauge  fields to generate extended gravity models going beyond the ordinary ones
in four dimensions. Omitting the scenarios of  the scalar fields originated from string theory compactification
mechanisms, the scalar fields could play a relevant role in the construction of certain black
 hole models. Roughly speaking, we consider a model described by the following action
\begin{equation}
S = \int d^{4}x \sqrt{-g}(R - V(\varphi) - (\nabla\varphi)^{2} - K(\varphi)F^{2})
\end{equation}
where R denotes the Ricci scalar curvature. $F^{2}$ being $F_{\mu\nu}F^{\mu\nu}$ is orginated from the Maxwell
gauge field. $\varphi$ is the stringy dilaton associated with a scalar potential V($\varphi$) and K($\varphi$) represents the stringy coupling function. A close examination shows that the coupling function
and the scalar potential could provide  certain  classes of black hole solutions. In the present work,
we consider a solution associated with the following functions 
\begin{eqnarray}
V(\varphi)&=&\frac{\lambda}{3} (e^{2\varphi} + 4 + e^{-2\varphi})\\
K(\varphi)& = &\frac{2e^{2\varphi}}{-2\gamma + \beta + \beta e^{2\varphi}}
\end{eqnarray}
where $\lambda$ can identified with  the cosmological constant. $\gamma$ and $\beta$ are extra parameters being
dimensionless providing an extended moduli space or parameters that constraint the EMS theory. It has been shown that an  exact black hole solution in EMS theory could be obtained by taking  $\gamma = -1$ \cite{Yu}. Moreover,  the EMD theory can be recovered by taking  $\beta=0$  \cite{Yu}. Considering the above function choice, one can obtain the rotating cosmological black hole metric in  EMS gravity theory by applying the Newman-Janis algorithm without complexification \cite{Ainou}.  In this way, the resulting metric  can be expressed as follows 
\begin{eqnarray}
ds^2&=& -\left( 1 - \frac{\sigma(r)}{H(r)}\right) dt^2 -\frac{2a^2\sigma(r)}{H(r)}  sin^2 \theta dt d\phi + \left( h(r) + a^2 +\frac{a^2\sigma(r)sin^2 \theta }{H(r)} \right) sin^2 \theta d\phi^2\notag\\
& +&
\frac{H(r)}{\triangle(r)} dr^2 + H(r) d\theta^2
\end{eqnarray}

where $a$ is the rotating spin parameter. The metric functions are given by
\begin{eqnarray}
H(r) &= &h(r) + a^2 cos^2 \theta\\
\triangle(r) &=& f(r)h(r) + a^2 \\
\sigma(r)&=& h(r)(1-f(r))
\end{eqnarray}
where one has used
\begin{equation}
f(r) = 1 - \frac{2M}{r} + \frac{\beta Q^2}{h(r)} - \frac{1}{3}\lambda h(r), \qquad
h(r) = r \left(1 -\frac{Q^2}{M r}\right).
\end{equation}

In this equation, $M$ and $Q$ are the mass and the charge of the black hole, respectively.
In what follows, we would like to investigate the extended moduli space effect on the
optical behaviors of this class of black holes. Precisely,    we approach the  $\beta$ contribution effects on   the associated optical behaviors.
\section{Constrained shadows of rotating  EMS black holes}
Optical aspects of black holes  have been largely
investigated using analytical and numerical approaches. These black hole activities have been
supported and encouraged by the observational results brought either by EHT collaborations
or by the gravitational wave detections \cite{EventHorizon}. A close inspection shows that two relevant
concepts have been studied being the shadow and the deflection angle of the light rays near
to black holes \cite{Chakhchi,Gogoi}. In this section, we reconsider the study of such optical quantities
for AdS black holes in EMS in terms of $\beta$ contributions.  Precisely, we provide constraints on the studied black holes
from the observational EHT data. These constraints will be exploited in the next section where a discussion on the deflection angle will be provided. 

\subsection{Shadow behaviors}
\begin{figure}[!ht]
		\begin{center}
		\centering
			\begin{tabbing}
			\centering
			\hspace{6.cm}\=\kill
			\includegraphics[scale=0.55]{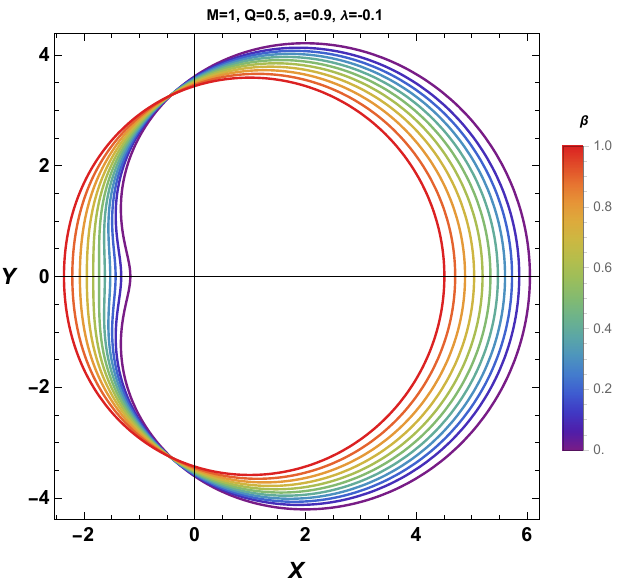} 
	\hspace{0.1cm}		\includegraphics[scale=0.55]{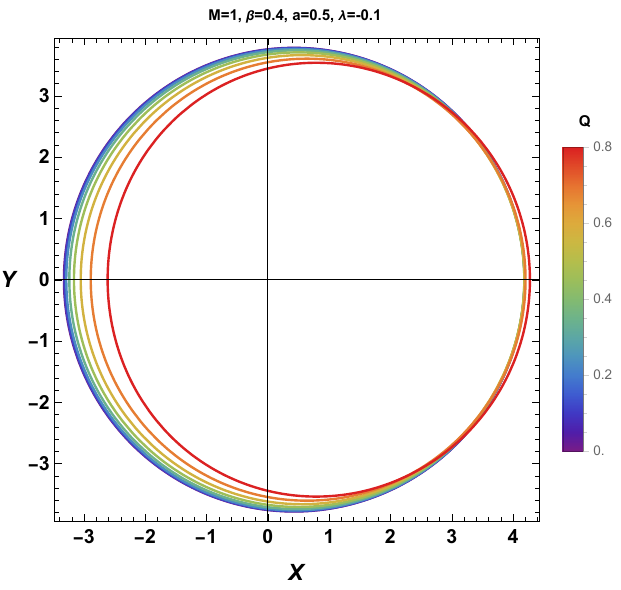}\hspace{0.1cm}	\includegraphics[scale=0.55]{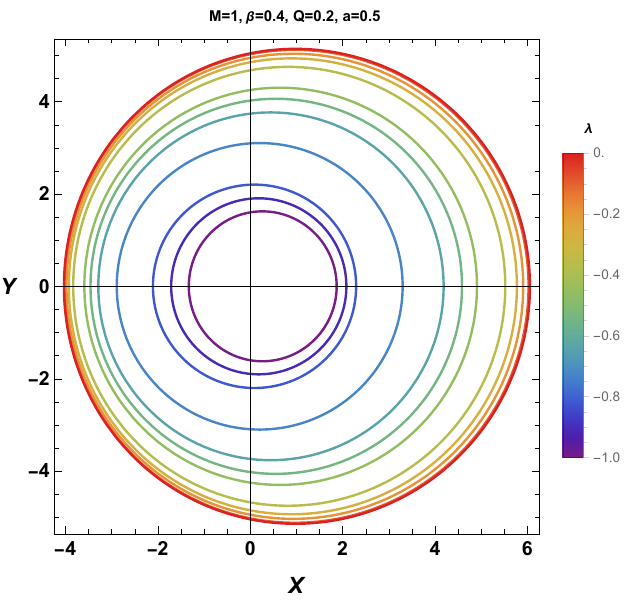}\\ 
	
		   \end{tabbing}
\caption{ \it \footnotesize  Shadow behaviors of  rotating AdS  EMS black holes in terms of the charge, the cosmological constant and the stringy coupling  parameter $\beta$.}
\label{rsa}
\end{center}
\end{figure}

In this part, we examine  the shadows of rotating  AdS black holes in EMS theory. It is recalled that the shadow of a black hole
is defined as the apparent boundary or the critical curve which appears when light rays
asymptotically approach an unstable circular orbit called photon sphere and return towards
the observer. This information is encoded on the null geodesics around black holes.
To approach the shadow optical aspects of the rotating black holes, we should establish
certain relations using the Hamilton-Jacobi factorization based on the Carter method \cite{A81}. Concretely, for such black hole solutions, the four equations of motion are given by
\begin{align}
H  \dot{t}& = \frac{h+a^2}{\Delta}\left[ E\left( h+a^2\right) -a L\right] +a\left[  L-aE\sin^2\theta \right]  \\
(H \dot{r})^2 &=\mathcal{R}(r) \\
( H\dot{ \theta})^2 & =\Theta(\theta)\\
H  \dot{\phi} &= \left[ L \csc ^2 \theta-aE\right] +\frac{a}{\Delta} \left[ E\left(h+a^2\right) -aL\right],
\end{align}
where $E$ and $L$ are the energy and the angular momentum of the light rays, respectively. $ \mathcal{R}(r)$  and $\Theta(\theta) $ functions  are expressed as follows
 \begin{align}
\mathcal{R}(r) &= \left[ E \left( h + a^2 \right) - a L \right]^2 - \Delta \left[ \mathcal{C} + \left( L - a E \right)^2 \right], \\
\Theta(\theta) &= \mathcal{C} - \left( L \csc \theta - a E \sin \theta \right)^2 + \left( L - a E \right)^2
\end{align}
where $\mathcal{C}$ is the carter separation  parameter.
Solving the  unstable circular orbit equations, the two needed impact parameters   are obtained as follows 
\begin{align}
& \eta=\frac{\mathcal{C}}{E^2} =\frac{4 \Delta (r) h'(r) \left(a^2 h'(r)+h(r) \Delta '(r)\right)-4 \Delta (r)^2 h'(r)^2-h(r)^2 \Delta '(r)^2}{a^2 \Delta '(r)^2}\vert_{r=r_0},\\
& \xi=\frac{L}{E}=\frac{a^2 \Delta '(r)-2 \Delta (r) h'(r)+h(r) \Delta '(r)}{a \Delta '(r)}\vert_{r=r_0}.
\end{align}
For the rotating AdS EMS  black holes, the apparent shapes of the shadows   are depicted in Fig.(\ref{rsa}) by considering the following celestial
coordinates 

\begin{equation}
\begin{aligned}
X & =\lim _{r_{ob } \rightarrow+\infty}\left(-r_{ob }^2 \sin \theta_{ob } \frac{d \phi}{d r}\right) \\
Y & =\lim _{r_{ob } \rightarrow+\infty}\left(r_{ob }^2 \frac{d \theta}{d r}\right),
\end{aligned}
\end{equation}
where $r_{ob }$ is the distance of the observer from the black hole.  It is denoted that  $\theta_{ob }$ indicates the angle of
inclination between the line of the observer and the axis of rotation of the black hole.

This figure provides  the shadow configurations for certain black hole parameters.  It  is worth noting that the deformation of the black hole shadow from the circular apparent is linked  to the no vanishing value of the rotating parameter. However, this effect is impacted by the contribution of the $\beta$ parameter.  As shown in the first panel of Fig.(\ref{rsa}), increasing the $\beta$ value from zero to one induces a gradual transition from a cardioid-like configuration to a D-shaped form, with  a reduced deformation and a smaller overall size. For the behavior of the shadow against the charge variation, it is clear from the second panel that the charge contributes as a  reducing parameter of the size  matching  with the previous study of the charged black holes. The point in this case is that the coefficient of the size deduction is more important for large value of the $\beta$ parameter. Interestingly, moving to the third panel of the  Fig.(\ref{rsa}), the behavior of the cosmological constant is amazingly  inverted. Taking large negative values gives small shadow configurations.
\subsection{Contact with empirical data via shadows}
To approach an alignment between rational predictions and empirical data,
the following subsection  presents an analysis of the shadow cast by rotating
cosmological EMS black holes, incorporating relevant data from the EHT.
Specifically, we exploit the EHT observational results for M87$^*$ and SgrA$^*$ to
establish constraints on the parameters of these black holes \cite{EventHorizon,Chakhchi,Gogoi}. Using the
fractional deviation from the Schwarzschild black hole shadow diameter, defined
as
\begin{equation}
\delta = \frac{R_{sh}}{r_{sh}},
\end{equation}

it has been demonstrated that these constraints can be imposed on the  black hole
parameters through the dimensionless quantity $R_s/M$. Indeed, the $(1-\sigma)$ and $(2-\sigma)$
measurements derived from the observational data are provided in Tab.(\ref{t1}).
\begin{table}[h!]
\centering
\begin{tabular}{|c|c|c|c|}
\hline
\textbf{Black Hole} & \textbf{Deviation (\(\delta\))} & \textbf{1-\(\sigma\) Bounds} & \textbf{2-\(\sigma\) Bounds} \\ 
\hline
M87$^*$ (EHT) & $-0.01^{+0.17}_{-0.17}$ & $4.26 \leq \frac{R_s}{M} \leq 6.03$ & $3.38 \leq \frac{R_s}{M} \leq 6.91$ \\ 
\hline
Sgr A$^*$ (EHT$_{\text{VLTI}}$) & $-0.08^{+0.09}_{-0.09}$ & $4.31 \leq \frac{R_s}{M} \leq 5.25$ & $3.85 \leq \frac{R_s}{M} \leq 5.72$ \\ 
\hline
Sgr A$^*$ (EHT$_{\text{Keck}}$) & $-0.04^{+0.09}_{-0.10}$ & $4.47 \leq \frac{R_s}{M} \leq 5.46$ & $3.95 \leq \frac{R_s}{M} \leq 5.92$ \\ 
\hline
\end{tabular}
\caption{Estimates and bounds for M87$^*$ and Sgr A$^*$ black holes.}
\label{t1}
\end{table}

\begin{figure}[!ht]
		\begin{center}
		\centering
			\begin{tabbing}
			\centering
			\hspace{6.cm}\=\kill
			\includegraphics[scale=0.53]{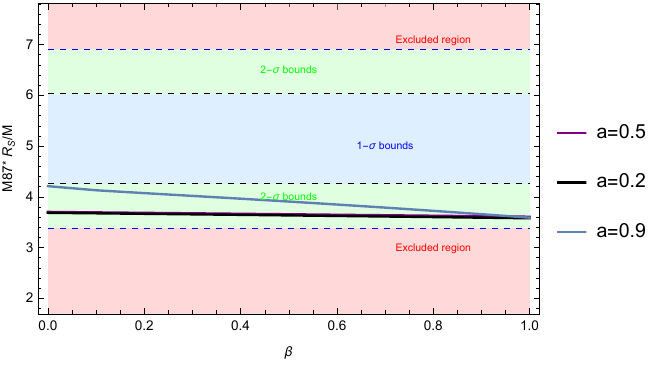} 
	\hspace{0.1cm}		\includegraphics[scale=0.53]{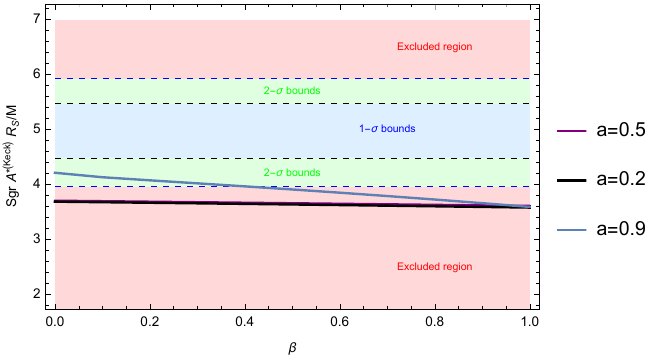}\hspace{0.1cm}	\includegraphics[scale=0.53]{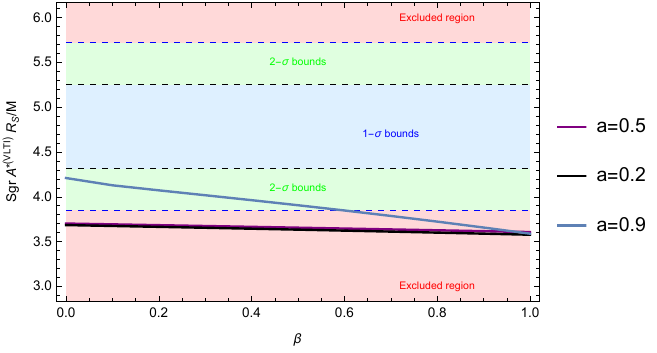}\\ 
	
		   \end{tabbing}
\caption{ \it \footnotesize Regions of cosmological rotating EMS black holes that are $(1-\sigma)$ and $(2-\sigma)$ consistent or inconsistent with the EHT data, as a function of the $\beta$ parameter. The plot shows the ratio $R_s/M$ for three values of the rotation parameter, with the charge $Q=0.2$, $\lambda=-0.1$ and the  mass $M=1$.}
\label{Cs1}
\end{center}
\end{figure}

\begin{figure}[!ht]
		\begin{center}
		\centering
			\begin{tabbing}
			\centering
			\hspace{6.cm}\=\kill
			\includegraphics[scale=0.51]{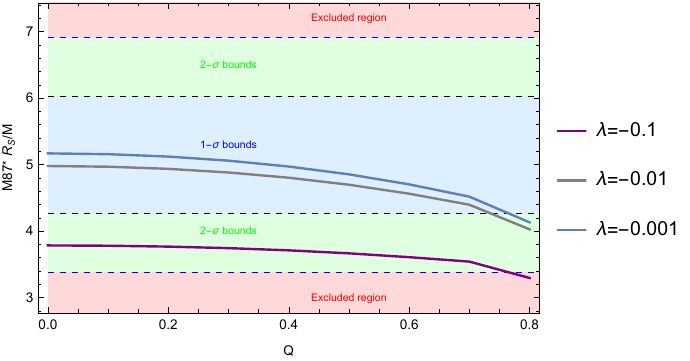} 
	\hspace{0.1cm}		\includegraphics[scale=0.51]{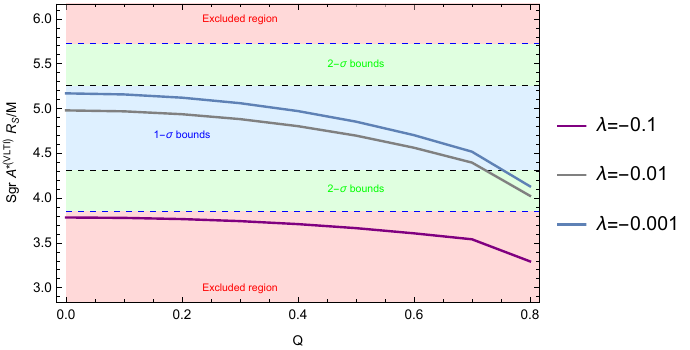}\hspace{0.1cm}	\includegraphics[scale=0.51]{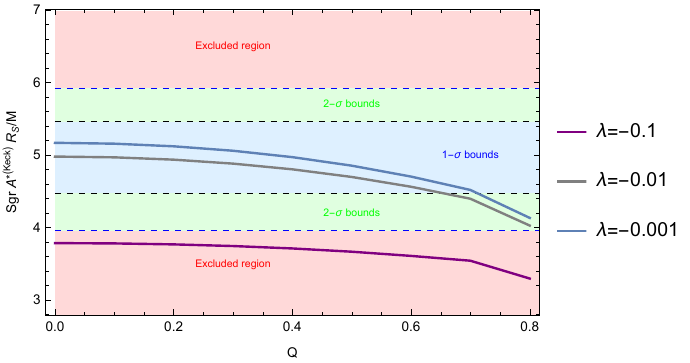}\\ 
	
		   \end{tabbing}
\caption{ \it \footnotesize  Regions of cosmological rotating EMS black holes that are $(1-\sigma)$ and $(2-\sigma)$ consistent or inconsistent with the EHT  empirical data, as a function of the charge. The plot shows the ratio $R_s/M$ for three values of the cosmological constant, with  $\beta=0.4$, $a=0.5$ and $M=1$.}
\label{Cs2}
\end{center}
\end{figure}

In Fig.(\ref{Cs1}), we illustrate the regions of cosmological rotating EMS AdS black holes that are
$(1 -\sigma)$ and $(2 -\sigma)$ consistent or inconsistent with the EHT data. These regions are shown
by varying the $\beta$ parameter for three different values of the rotation parameter.
 Based on the figure, achieving
consistency with the EHT data  requires a  small  value of the stringy  parameter  $\beta$  for  relevant  values of the rotating parameter. However, there is  a non  $(1 -\sigma)$ consistency. Fixing the rotating parameter to 0.9,  the shadow
radii align closely with the observational $(2 - \sigma)$ constraints from EHT for M87* and Sgr A* for all the $\beta$ values less than 0.4. We move now to unveil the effect of the charge and the cosmological constant on the studied black holes.  Fig.(\ref{Cs2}) shows that  cosmological rotating EMS AdS black holes  
$(1 -\sigma)$ and $(2 -\sigma)$  are  consistent with the EHT data for very small values of the cosmological constant. This confirms the previous study of the $\lambda$ effect on EMD black holes  \cite{}. 

\section{ Deflection angle behaviors}
\begin{figure}[!ht]
		\begin{center}
		\centering
			\begin{tabbing}
			\centering
			\hspace{6.cm}\=\kill
			\includegraphics[scale=0.53]{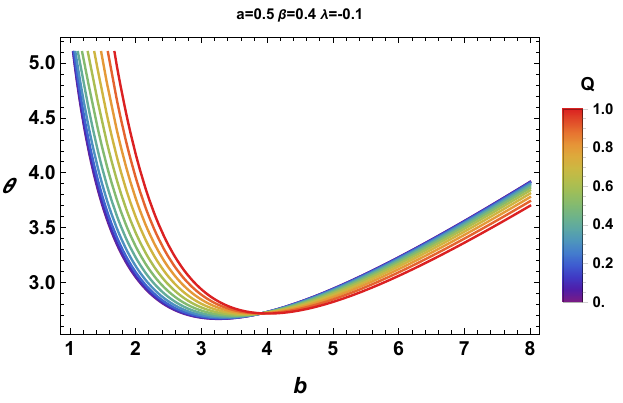} 
	\hspace{0.1cm}		\includegraphics[scale=0.53]{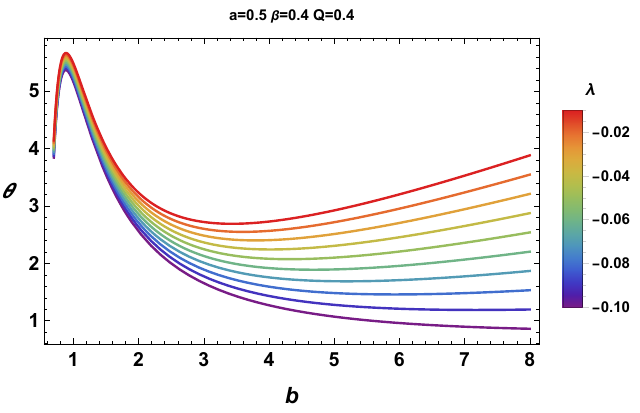}\hspace{0.1cm}	\includegraphics[scale=0.4]{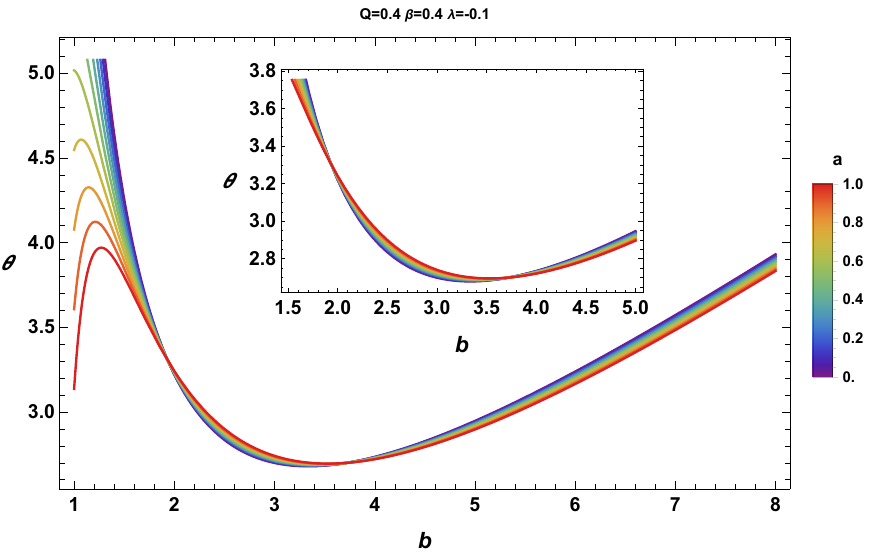}\\ 
	
		   \end{tabbing}
\caption{ \it \footnotesize Behaviors of the deflection angles as a function of the impact parameter by varying  the charge, the cosmological constant and the rotating parameters.}
\label{Cs5}
\end{center}
\end{figure}

In this section,  we approach   the second optical quantity. It turns out that the essence of gravitational
lensing lies in the bending of light due to gravitational fields, as predicted by Einstein
general relativity, especially in the weak field limit. This phenomenon occurs
when the light passes near massive objects like planets, black holes, or  other matter. The weak
deflection,  being a consequence of this gravitational lensing, is a valuable tool in detecting non-trivial 
matter filaments. Understanding the weak deflection is crucial as it provides
insights into the large-scale structure of the universe. To calculate the deflection angle,
we will adopt the approach based on the Gauss-Bonnet theorem. The latter  is one of the important
method to calculate the weak deflection angle using optical geometry proposed by Gibbons
and Werner \cite{A126,A127}. This theorem establishes a connection between the intrinsic differential
geometry of a surface and its topology. Considering the observer and the source at finite
distance in the equatorial plane, the deflection angle can be expressed as
\begin{equation}
\Theta = \Psi_R - \Psi_S + \phi_{SR}
\end{equation}
where $\Psi_R $ and $ \Psi_S$ represent the angles between the light rays and the radial
direction at the positions of the observer and the source, respectively. The angle
$\Psi_{SR}$ denotes the longitudinal separation between these positions, as described in
\cite{A118,A124,A117}. To calculate these optical quantities, we can employ the algorithm developed in \cite{A126,A127,A118,A124,A117}. Roughly, considering the impact parameter  $b$ to be the ratio between the energy and  the angular momentum of the light rays and taking the calculation to be in the order  $\mathcal{O}(M^1,\lambda^1,Q^2,a^1,\beta^1)$, we can  reveal  that  the light deflection angle can be factorized
using two  contributions   associated with  the absence and the presence of  the stringy  coupling parameter $\beta$.  Instead of providing very large computations, we give only the essential ones. Indeed,   the deflection angle  can be split into parts as follows 
\begin{equation}
\Theta (\beta) =  \Theta_{EMD}(M,a,Q,\lambda)+  \Theta_\beta(M,a,Q,\lambda) \beta 
\end{equation}
where $\Theta_{EMD}(M,a,Q,\lambda)$ is the deflection angle of  the Einstein-Maxwell-dilaton black holes  given by 
{\small
\begin{eqnarray}
\Theta_{EMD}(M,a,Q,\lambda) &=& \frac{4 M}{b}+\frac{16 M Q^2}{b^3} -\frac{b \lambda }{6}  \left(\frac{1}{{u_R}}+\frac{1}{{u_S}}\right)-b \lambda  M+\frac{8 \lambda  M Q^2}{b}+\frac{b \lambda  Q^2}{6 M}+\frac{2 a M}{b^2}+\frac{4 a \lambda  M}{3}\notag \\&-&\frac{24 a M Q^2}{b^4} -\frac{12 a \lambda  M Q^2}{b^2}
\end{eqnarray}}
where  $u_S$ and $u_R$ are the inverse of the distances from the black hole to the
source and the observer, respectively.  $\Theta_\beta(M,a,Q,\lambda)$  represents the  $\beta$ contribution coefficient  being expressed as follows
{\small
\begin{eqnarray}
\Theta_\beta(M,a,Q,\lambda)&=&\frac{16 M Q^2}{b^3}+\frac{8 \lambda  M Q^2}{b}-\frac{48 a M Q^2}{b^4}-\frac{24 a \lambda  M Q^2}{b^2}.
\label{bb}
\end{eqnarray}}
This extra  term  shows that the  deflection angle  depends  on the stringy parameter.
Putting $\beta=0$, we  recover  the expression of the deflection angle of the light rays around EMD black holes  found recently in \cite{hemd}. Moreover,  the  parameter $\beta$ contributes   linearly to  the light deviation angle.  This contribution   depends on the other black hole parameter.  As the above discussion,   the stringy parameter  $\beta$  could be considered as  an increasing or a decreasing parameter  of the deflection angle. This is controlled by  the sign of the coefficient function $\Theta_\beta(M,a,Q,\lambda)$.
As the compatibility with empirical results necessitates a small value of the stringy  parameter $\beta$, we  consider $\beta=0.4$  to illustrate such behaviors.  In Fig.(\ref{Cs5}),  we plot the deflection angle against the other parameters of the black hole to discover how the $\beta$ contribution  changes  their effects.

In Fig.(\ref{Cs5}), we observe no  difference in the behavior of the deflection angle with respect to variations in the charge and the cosmological parameter  compared with the ordinary black holes. This  angle continues to decrease as a function of the cosmological constant, and the concave trend is attributed to the contribution of this cosmological parameter. Varying the  charge, the deflection angle initially increases with the charge up to a critical point, beyond which the trend changes, with greater deviations occurring at smaller charge values. A notable difference is observed by varying  the rotation parameter $a$. Unexpectedly, the behavior of the deflection angle against this rotation parameter differs from other ordinary cases and does not occur when we take $\beta=0$. The rotation parameter initially decreases the deflection angle for small values of the impact parameter. However, within a specific range of the  impact parameter values, the rotation parameter begins to act as an increasing factor for the deflection angle, before ultimately reverting to a decreasing influence for larger impact parameter values.

\begin{figure}[!ht]
		\begin{center}
			
	\includegraphics[scale=0.55]{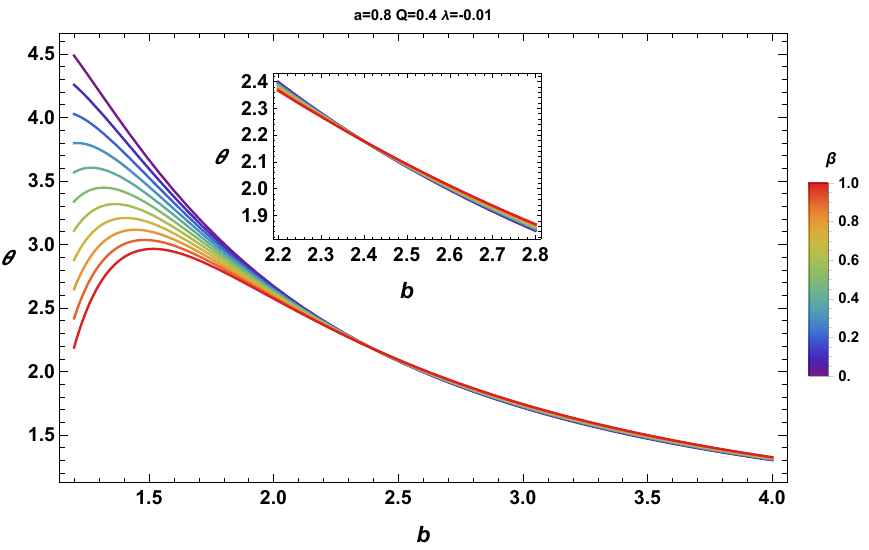}\hspace{0.1cm}

\caption{ \it \footnotesize  Variation of the deflection angle in terms of the rotating parameter for different values of $\beta$ parameter.}
\label{Cs15}
\end{center}
\end{figure}

To examine the contribution of 
$\beta$ more closely, in Fig.(\ref{Cs15}), we fix the black hole parameter to be 
$(1 -\sigma)$ and $(2 -\sigma)$ consistent with the Event Horizon Telescope (EHT) data and we plot the deflection angle as a function of the impact parameter by varying 
$\beta$ from 0 to 1. The first observation from the figure is that the concavity is almost eliminated, likely due to the small value of the cosmological constant discussed earlier. We also see that the deflection angle decreases as a function of the rotation parameter, resembling the behavior of the Schwarzschild deflection angle.  Moreover, the behavior of the deflection angle with respect to the 
$\beta$ parameter is controlled by the impact parameter values, as illustrated in the figure and supported by Eq.(\ref{bb}). For small values of the impact parameter, the deflection angle decreases by  increasing  the 
$\beta$  parameter.  For larger values of the impact parameter,  however, this behavior reverses. This is expected from the metric expression for this type of black hole, where the coupling between charge and the stringy  parameter  $\beta$   suggests that they will contribute in a similar manner.
\section{Conclusion}
Motivated by EHT results, we have  investigated certain  optical properties of four
dimensional black holes with AdS geometries in EMS gravity. In particular, we have  approached   first  the shadows of rotating   black holes in EMS gravity by introducing a stringy
parameter $\beta$. Using Newman-Janis method and the Hamilton-Jacobi algorithm,  we have
obtained various geometries,  including D and cardioid ones, which depend on the rotating parameter and the stringy gravity
parameter $\beta$. In particular, we have shown that such a parameter can modify the previous
findings. Sending this stringy  parameter to zero, we have recovered the usual black hole shadows
like the ones of ordinary solutions. Exploiting the EHT data, we have presented predictions
for the stringy gravity parameter which could be exploited in string theory compactifications. Among others, the obtained shadow geometries have shown certain matchings with
EHT collaboration data by imposing some constraints on the involved black hole parameters. Then, we have studied light behaviors near  to AdS black holes. In particular, we have
computed and analyzed the deflection angle of light rays near to the studied black holes.  Along others, we have shown  that such an optical quantity   can be factorized
using two  contributions   associated with  the absence and the presence of  the string coupling parameter $\beta$.   Enlarging  the  black hole moduli space, the deflection angle 
could be split into several  stringy contributions such as $
\Theta (\beta) =  \Theta_{EMD}(M,a,Q,\lambda)+  \sum\limits_i \Theta_{\beta_i}(M,a,Q,\lambda) \beta_i$ where $ \beta_i$  could be identified with certain stringy  parameters   This could
open new road  to explore other physical  constraints on the optical quantities using string theory compactification scenarios.

This work raises certain issues. Motivated by recent investigations
of quantum information concepts in LHC activities, it could be interesting to implement the
shadow properties of black holes in this scene. This could open new gates to understand the
underlying physics. We hope to address such questions in future studies.

\section*{Acknowledgements}  The author  would like to thank Adil Belhaj and Yassine Hassouni for  collaboration, encouragement, and  scientific support. She is also grateful    for their valuable comments and insightful discussions.


\begin{thebibliography}{10}

\bibitem{A56}
K. Akiyama \textit{et al.}, \textit{First M87 Event Horizon Telescope Results. I. The Shadow of the Supermassive Black Hole}, Astrophys. J. \textbf{875}, 01 (2019) L1, {\tt arXiv:1906.11238}.

\bibitem{B10}
K. Akiyama \textit{et al.}, {\it First M87 Event Horizon Telescope Results. VIII. Magnetic Field Structure near The Event Horizon}, Astrophys. J. Lett. \textbf{910}, 1 (2021)  L13, {\tt arXiv:2105.01173}.

\bibitem{A57} 
K. Akiyama \textit{et al.}\textit{First M87 Event Horizon Telescope Results. IV. Imaging the Central Supermassive Black Hole}, Astrophys. J. Lett. \textbf{875},  01 (2019) L4.

\bibitem{A58}
K. Akiyama \textit{et al.}, {\it First Sagittarius A* Event Horizon Telescope Results. I. The Shadow of the Supermassive Black Hole in the Center of the Milky Way}, Astrophys. J. Lett. \textbf{930},  2 (2022)  L12, {\tt arXiv:1906.11241}.

\bibitem{A59}
K. Akiyama \textit{et al.}, {\it First Sagittarius A* Event Horizon Telescope Results. II. EHT and Multiwavelength Observations, Data Processing, and Calibration}, Astrophys. J. Lett. \textbf{930}, 2 (2022) L13.


\bibitem{A60}
K. Akiyama \textit{et al.}, {\it First Sagittarius A* Event Horizon Telescope Results. III. Imaging of the Galactic Center Supermassive Black Hole}, Astrophys. J. Lett. \textbf{930}, 2 (2022) L14.

\bibitem{LIGs}
B. Abbott and al,
\textit{Binary Black Hole Population Properties Inferred from the First and Second Observing Runs of Advanced LIGO and Advanced Virgo},
Astrophys. J. Lett. \textbf{882} (2019) no.2, L24.

\bibitem{BGRA}
     B. Abbott and al., \textit{Observation of Gravitational Waves from a Binary Black Hole Merger},
Phys.\ Rev.\ Lett.\ {\bf 116} (6) (2016) 061102, {\tt arXiv:1602.03837.}

\bibitem{Jaf}
K.~Jafarzade, B.~Eslam Panah and M.~E.~Rodrigues,
\textit{Thermodynamics and optical properties of phantom AdS black holes in massive gravity},
Class. Quant. Grav. \textbf{41} (2024) no.6, 065007.

\bibitem{Yanxx}
S.~F.~Yan, C.~Li, L.~Xue, X.~Ren, Y.~F.~Cai, D.~A.~Easson, Y.~F.~Yuan and H.~Zhao,
\textit{Testing the equivalence principle via the shadow of black holes},
Phys. Rev. Res. \textbf{2} (2020) no.2, 023164

\bibitem{A154A}
A.~Belhaj, H.~Belmahi, M.~Benali, M.~Oualaid and M.~B.~Sedra, {\it Light Trajectories and Thermal Shadows casted by Black Holes in a Cavity},     JCAP \textbf{11} (2023) 094,  {\tt arXiv:2206.00615}. 

\bibitem{A65}
B. P. Singh,  S. G.  Ghosh, \textit{Shadow of Schwarzschild–Tangherlini black holes}, Annals of Physics {\bf 395} (2018) 127, {\tt arXiv:1707.07125}.

\bibitem{A66}
V. Perlick, O. Y. Tsupko,  G. S. Bisnovatyi-Kogan, \textit{Black hole shadow in an expanding universe with a cosmological constant}, Phys. Rev. D \textbf{97}, 10 (2018) 104062, {\tt arXiv:1804.04898}.

\bibitem{A68}
R. Shaikh, P. Kocherlakota, R. Narayan,  P. S. Joshi, \textit{Shadows of spherically symmetric black holes and naked singularities}, Mon. Not. Roy. Astron. Soc. \textbf{482}, 1 (2019) 52,  {\tt  arXiv: 1802.08060}.

\bibitem{A69}
T. Zhu, Q. Wu, M. Jamil, K. Jusufi, \textit{Shadows and deflection angle of charged and slowly rotating black holes in Einstein-{\AE}ther theory}, Phys. Rev. D {\bf100} (2019) 044055, {\tt arXiv:1906.05673}.


\bibitem{A74}
A. Belhaj, M. Benali, A. El Balali, H. El Moumni and S. E. Ennadifi, {\it Deflection angle and shadow behaviors of quintessential black holes in arbitrary dimensions}, Class. Quant. Grav. \textbf{37}, 21 (2020) 215004, {\tt arXiv:2006.01078}. 



\bibitem{A78}
R. Uniyal, N. Chandrachani Devi, H. Nandan and K. D. Purohit, {\it Geodesic Motion in Schwarzschild Spacetime Surrounded by Quintessence}, Gen. Rel. Grav. \textbf{47}, 2 (2015) 16, {\tt arXiv:1406.3931}.

\bibitem{A81}
S. W. Wei, Y. C. Zou, Y. X. Liu,  R. B. Mann, \textit{Curvature radius and Kerr black hole shadow}, JCAP \textbf{08} (2019) 030,  {\tt arXiv:1904.07710}.



 \bibitem{MHER}
Z.~S.~Moreira, C.~A.~R.~Herdeiro and L.~C.~B.~Crispino,
Twisting shadows: Light rings, lensing, and shadows of black holes in swirling universes,
Phys. Rev. D \textbf{109} (2024) no.10, 104020.
\bibitem{HERDE}
L.~K.~Wong, C.~A.~R.~Herdeiro and E.~Radu, Constraining spontaneous black hole scalarization in scalar-tensor-Gauss-Bonnet theories with current gravitational-wave data,
Phys. Rev. D \textbf{106} (2022) no.2, 024008

\bibitem{Novo}
J.~P.~A.~Novo, P.~V.~P.~Cunha and C.~A.~R.~Herdeiro, Hypershadows of higher dimensional black objects: a case study of cohomogeneity-one d=5 Myers-Perry,
[arXiv:2410.05390 [gr-qc]].



\bibitem{SV1}
S.~Vagnozzi, R.~Roy, Y.~D.~Tsai, L.~Visinelli, M.~Afrin, A.~Allahyari, P.~Bambhaniya, D.~Dey, S.~G.~Ghosh and P.~S.~Joshi, \textit{et al.}
Horizon-scale tests of gravity theories and fundamental physics from the Event Horizon Telescope image of Sagittarius A,
Class. Quant. Grav. \textbf{40} (2023) no.16, 165007.

{A79,A159,SV2}
\bibitem{A79}
A. Abdujabbarov, M. Amir, B. Ahmedov and S. G. Ghosh,{\it Shadow of rotating regular black holes}, Phys. Rev. D \textbf{93} (2016) 104004, {\tt arXiv:1604.03809}. 


\bibitem{A159}
A.~Belhaj, H.~Belmahi, M.~Benali, H.~El Moumni, M.~A.~Essebani and M.~B.~Sedra, {\it Optical shadows of rotating Bardeen-AdS black holes}, Mod. Phys. Lett. A \textbf{37}, 06  (2022) 2250032, {\tt arXiv:2202.10892}. 

\bibitem{SV2}
D.~Pedrotti and S.~Vagnozzi, Quasinormal modes-shadow correspondence for rotating regular black holes,
Phys. Rev. D \textbf{110} (2024) no.8, 084075.
\bibitem{A77}
O. Pedraza, L. A. L\'opez, R. Arceo and I. Cabrera-Munguia, {\it Geodesics of Hayward black hole surrounded by quintessence}, Gen. Rel. Grav. \textbf{53} (2021) 24, {\tt arXiv:2008.00061}.

\bibitem{sbrh}
A.~Belhaj, H.~Belmahi, M.~Benali, Y.~Hassouni and M.~B.~Sedra,
 {\it Optical behaviors of black holes in Starobinsky\textendash{}Bel\textendash{}Robinson gravity},
Gen. Rel. Grav. \textbf{55}, no.10, 110 (2023).
\bibitem{Lekbi}
H.~Lekbich, N.~Parbin, D.~J.~Gogoi, A.~E.~Boukili and M.~B.~Sedra, The optical features of noncommutative charged 4D-AdS-Einstein\textendash{}Gauss\textendash{}Bonnet black hole: shadow and deflection angle,
Eur. Phys. J. C \textbf{84} (2024)  350.
\bibitem{YASSS}
A.~Al-Badawi, Y.~Sekhmani, J.~Rayimbaev and R.~Myrzakulov, Shadows and quasinormal modes of black holes in the Einstein-SU(N) nonlinear sigma model,
Int. J. Mod. Phys. D \textbf{33} (2024) no.12, 2450043.
\bibitem{Tsu}
N.~Tsukamoto, Z.~Li and C.~Bambi, Constraining the spin and the deformation parameters from the black hole shadow,
JCAP \textbf{06} (2014) 043.
\bibitem{BBAM}
M.~Benali and A.~E.~Balali, Rotating reduced Kiselev black holes: Shadows, Energy emission and Deflection of light,
{\tt arXiv:2406.00788.} 
\bibitem{AMM}
A.~E.~Balali, M.~Benali and M.~Oualaid, Deflection angle and shadow of slowly rotating black holes in galactic nuclei,
Gen. Rel. Grav. \textbf{56} (2024) no.2, 21.

\bibitem{A95}
A. Belhaj, M. Benali, A. El Balali, W. El Hadri, H. El Moumni and E. Torrente-Lujan, {\it Black hole shadows in M-theory scenarios}, Int. J. Mod. Phys. D \textbf{30} (2021) 2150026, {\tt arXiv:2008.09908}.

\bibitem{A89}
A. Belhaj, A. El Balali, W. El Hadri, Y. Hassouni, E. Torrente-Lujan, {\it Phase transition and shadow behaviors of quintessential black holes in M-theory/superstring inspired models},  Int.J.Mod.Phys. A {\bf 36} (2021) 2150057.

\bibitem{A96}
 A. Belhaj, H. Belmahi,  M. Benali, W. El Hadri, H. El Moumni, E. Torrente-Lujan,  \textit{Shadows of 5D Black Holes from string theory},  Phys. Lett.B \textbf{812} (2021) 136025, {\tt arXiv:2008.13478}.
\bibitem{HRC}
A.~Belhaj, H.~Belmahi, M.~Benali and H.~Moumni El,
{\it  Light deflection by rotating regular black holes with a cosmological constant},
Chin. J. Phys. \textbf{80} (2022)229-238.
\bibitem{HADS}
 A.~Belhaj, H.~Belmahi and M.~Benali,
{\it Deflection light behaviors by AdS black holes},
Gen. Rel. Grav. \textbf{54}, no.1, (2022) 4.
\bibitem{HMTD}
 N. Askour, A. Belhaj, H. Belmahi, M. Benali, H. El Moumni and Y. Sekhmani,  {\it Deflection angle and light ray
trajectories near M-theory black holes}, IJGMMP, 191 (2023) 01, 2450019.


\bibitem{Hirschmann}
E.~W.~Hirschmann, L.~Lehner, S.~L.~Liebling and C.~Palenzuela,
\textit{Black Hole Dynamics in Einstein-Maxwell-Dilaton Theory,}
Phys. Rev. D \textbf{97} (2018) no.6, 064032.
\bibitem{Yu}
S.~Yu, J.~Qiu and C.~Gao,
\textit{Constructing black holes in Einstein\textendash{}Maxwell-scalar theory,}
Class. Quant. Grav. \textbf{38} (2021) no.10, 105006.
\bibitem{Al}
\textit{A.~Al-Badawi, M.~Alloqulov, S.~Shaymatov and B.~Ahmedov,
Shadows and weak gravitational lensing for black holes within Einstein-Maxwell-scalar theory,}
Chin. Phys. C \textbf{48} (2024) no.9, 095105.
\bibitem{Charmousis}
C.~Charmousis, B.~Gouteraux and J.~Soda,
\textit{Einstein-Maxwell-Dilaton theories with a Liouville potential,}
Phys. Rev. D \textbf{80} (2009) 024028.

\bibitem{Lu}
H.~L\"u, Z.~L.~Wang and Q.~Q.~Zhao,
\textit{Black Holes That Repel},
Phys. Rev. D \textbf{99} (2019) no.10, 101502.
\bibitem{hemd}
H.~Belmahi and A.~M.~Rbah,
{\it Optical Aspect of Cosmological Black Holes in Einstein-Maxwell-Dilaton Theory,}   {\tt  
arXiv:2409.08903.}
\bibitem{Ainou}
M.~Azreg-A\"\i{}nou,
\textit{Generating rotating regular black hole solutions without complexification,}
Phys. Rev. D \textbf{90} (2014) no.6, 064041.

\bibitem{EventHorizon}
P.~Kocherlakota \textit{et al.} [Event Horizon Telescope],
\textit{Constraints on black-hole charges with the 2017 EHT observations of M87*,}
Phys. Rev. D \textbf{103} (2021) no.10, 104047.
\bibitem{Gogoi}
D.~J.~Gogoi and S.~Ponglertsakul,
\textit{Constraints on quasinormal modes from black hole shadows in regular non-minimal Einstein Yang\textendash{}Mills gravity,}
Eur. Phys. J. C \textbf{84} (2024) no.6, 652.
\bibitem{Chakhchi}
L.~Chakhchi, H.~El Moumni and K.~Masmar,
\textit{Signatures of the accelerating black holes with a cosmological constant from the Sgr A\ensuremath{\star} and M87\ensuremath{\star} shadow prospects,}
Phys. Dark Univ. \textbf{44} (2024) 101501.

\bibitem{A126}
G. W. Gibbons and M. C. Werner, {\it Applications of the Gauss-Bonnet theorem to gravitational lensing}, Class. Quant. Grav. \textbf{25} (2008) 235009, {\tt arXiv:0807.0854}.

\bibitem{A127}
A. Ishihara, Y. Suzuki, T. Ono, T. Kitamura, H. Asada, {\it Gravitational bending angle of light for finite distance and the Gauss-Bonnet theorem}, Phys. Rev. D \textbf{94} (2016) 084015, {\tt arXiv:1604.08308}.

\bibitem{A118}
T. Ono, A. Ishihara,  H. Asada, \textit{Gravitomagnetic bending angle of light with finite-distance corrections in stationary axisymmetric spacetimes}, Phys. Rev. D \textbf{96} (2017) 104037, {\tt  arXiv:1704.05615}.


 \bibitem{A124}
K. S. Virbhadra, G. F. Ellis, \textit{Schwarzschild black hole lensing}, Phys. Rev. {\bf D62}, 8 (2000) 084003.
\bibitem{A117}
W. Javed, J. Abbas,   A. Övgün,  \textit{Deflection angle of photon from magnetized black hole and effect of nonlinear electrodynamics}, Eur. Phys. J. C \textbf{79}  (2019) 694, {\tt arXiv:1908.09632}.
\end{thebibliography}
\end{document}